\documentstyle[prc,aps,epsf]{revtex}
\draft
\preprint{\today}
\begin{document}

\title{Shape and blocking effects on odd-even mass differences and
rotational motion of nuclei}

\author{F.R. Xu,$^{1,}$\footnote{Permanent address:
Department of Technical Physics, Peking University, Beijing 100871, China}
R. Wyss,$^2$ and P.M. Walker$^1$ } 

\address{$^1$Department of Physics, University of Surrey,
         Guildford, Surrey GU2 5XH, England}
\address{$^2$Department of Physics, Royal Institute of Technology,
         Frescativ{\"a}gen 24, S-104 05 Stockholm, Sweden}

\maketitle 

\begin{abstract}
Nuclear shapes and odd-nucleon blockings strongly influence the
odd-even differences of nuclear masses. When such effects are
taken into account, the determination of the 
pairing strength is modified resulting in larger pair gaps.
The modified pairing strength leads
to an improved self-consistent description of moments of inertia 
and backbending frequencies, with no additional parameters.
\end{abstract}

 
Since the BCS theory was applied to atomic nuclei \cite{Bel59,Nil61},
pairing correlations have been crucial to the understanding of
many properties, such as binding energies, collective rotational
motion and quasiparticle excitation energies. 
The interaction strength, $G$, of the pairing force
is the key parameter that governs the properties of the short range
correlations.

The $G$ value is usually determined by fitting the BCS pairing gaps
($\Delta = G \sum_i U_i V_i$) of even-even nuclei
to experimental odd-even mass differences, $D_{\rm oe}$ (where $U_i$ and $V_i$
are the emptiness and occupation amplitudes of nucleon pairs)\cite{Nil61}.
However, when one calculates theoretical $D_{\rm oe}$
values (i.e. in the same manner as calculating experimental $D_{\rm oe}$
values, but with theoretical nuclear masses) with the $G$-value determined
according to the above prescription,
it turns out that they 
do not agree with experiment.
The theoretical values are 
systematically smaller than the pairing gaps,
at least for the rare-earth deformed nuclei described below.
In principle, experimental odd-even mass differences should be
compared with theoretical $D_{\rm oe}$ values.
Hence, the pairing strength $G$ should be adjusted to reproduce, at least
on average, theoretical odd-even masses and not pair gaps of even-even nuclei.

Although pairing correlations are dominantly responsible for
odd-even mass differences, there exist other non-negligible
effects, in view of the systematic differences mentioned 
above\cite{BM69}.
One important effect stems from the deformed mean field.
Due to the Kramers degeneracy of
single-particle levels, odd- and even-nucleon systems will
have different energies in a deformed field. The interplay between
pairing and this `mean field' effect
has been clarified in a recent work by Satula {\it et al.} \cite{Sat98}.
For light- and
medium-mass nuclei, it is comparable with the pairing 
contribution\cite{Sat98}.
Furthermore, when neighbouring nuclei (which are involved
in calculating odd-even mass differences) 
have different deformations, shape-changing effects will also play a role.
These two factors originate from the mean field and
we will refer to them in the following as {\sl shape} effects.

Another important influence is the blocking effect.
Experimental odd-even mass differences
contain the odd-nucleon blocking effect in adjacent
odd-particle systems, which is absent in even systems\cite{Nil61,Zen83}.
This effect can become significant, especially
when the densities of single-particle levels around the neutron and proton
Fermi surfaces are not very high.
Simple BCS calculations typically show that odd-nucleon blockings reduce
pairing gaps by more than 10\% for the nuclei in the rare-earth region.
Both the shape and blocking effects will influence the determination of
the pairing strengths.

Two very sensitive probes of pairing correlations, and therefore of the
pairing strengths, are 
moments of inertia (see e.g. \cite{Nil61,RBen86,Goe90})
and backbending (bandcrossing) frequencies \cite{RBen83,TBen89}.
States of high seniority may serve as another probe, and 
recent calculations of the energies of
multi-quasiparticle states show the need for 
adjustment of the pairing
strength\cite{Xu98}. Hence, the question arises as to whether
the pairing strength determined from odd-even mass differences
is consistent with the pairing strength used to calculate
moments of inertia or energies of high seniority states.
A consistent way to determine the
$G$-value is therefore an important issue for the quantitative
description of nuclear properties. In this paper, we 
show that when shape
and blocking effects are taking into account,
the pairing strength $G$ needs to be modified
in order to reproduce experimental $D_{\rm oe}$ values.
Such modifications result in an improved self-consistent
description for both moments of inertia and band-crossing frequencies.

In order to minimize the influences of the quantities that are not relevant
for our discussion, we use the five-point formula of Ref.\cite{Mol92}
to determine experimental $D_{\rm oe}$ values. For an even-even nucleus,
\begin{equation}
D_{\rm oe}=-{\frac{1}{8}}[M(N+2)-4M(N+1)+6M(N)-4M(N-1)+M(N-2)]\;,
\end{equation}
where $M(N)$ is the mass of an atom with neutron number, $N$
(or $Z$ for protons). The quantity, $D_{\rm oe}$, is calculated
along an isotopic (or isotonic) chain.
With Eq.(1) we investigate shape and blocking effects using the
deformed Woods-Saxon (WS) model \cite{Naz85,NAZ103}.
According to the Strutinsky energy theorem \cite{Bra72}, 
the total energy of a nucleus can be decomposed into
a macroscopic and microscopic part. The latter consists of
shell and pairing correction energies.
For the macroscopic energy, we employ the standard liquid-drop model
of Ref.\cite{Mye69}.
Pairing correlations are treated by a technique of 
approximate particle-number projection, known as the Lipkin-Nogami
(LN) method \cite{Pra73} which takes particle-number-fluctuation
effects into account by introducing an additional Lagrange multiplier,
$\lambda_2$. Both monopole and quadrupole pairings are included
for residual two-body interactions.

Nuclear shapes are determined by minimizing calculated
potential-energy-surface (PES) energies
in the quadrupole deformation ($\beta_2, \gamma$)
space with hexadecapole ($\beta_4$) variation. 
For well-deformed nuclei, pairing energies only weakly
influence equilibrium deformations\cite{Xu98,Bra72}.
Therefore, we use the monopole pairing strengths
obtained by the average gap method \cite{Mol92} to
determine nuclear deformations.
The quadrupole pairing strength is determined by restoring the local
Galilean invariance with respect to quadrupole
shape oscillations \cite{Sak90,Satu94}.
Whereas quadrupole pairing is essential for the proper description
of the moments of inertia \cite{Die84,Wys95}, its influence
on nuclear binding energies is negligible, since we
use the doubly-stretched quadrupole operators \cite{Satu94,Sak89}.

In the present work, we focus on well-deformed
rare-earth nuclei where an abundance of regular collective rotational
bands with backbending have been observed.
The shape effects, coming from the shell-correction and the
macroscopic deformation energies,
can be calculated using Eq.(1),
after determining the equilibrium deformations.
Our calculations show that
the shape effects are usually of the order of 100 to 200 keV
for a range of even-even Er, Yb, Hf and W isotopes.
If one neglects the changes of deformation,
the shape effect for a deformed even system ($N = 2 n$)
can be written as $\frac{1}{2}(e_{n+1}-e_n)$ for the three-point
formula \cite{Sat98} or $\frac{1}{4}(e_{n+1}-e_n)$ for the five-point
formula of  Eq.(1) (where $e_i$ is the single-particle energy).
Shape effects calculated from the above simple forms 
differ from those calculated according
to Eq.(1), when shape changes are included.
This implies that the polarization effects of the
odd nucleons have to be considered explicitly, as
is done in the present work.
In contrast to light nuclei \cite{Sat98}, the mean-field effects for
heavy nuclei are not so large due to the relatively close spacing of
the single-particle levels.

In the LN model (for the case of monopole pairing) the quantity,
$\Delta+\lambda_2$, is assumed to be identified \cite{Mol92} with the
odd-even mass difference, $D_{\rm oe}$, provided that other physical influences
(e.g. shape and blocking effects) are ignored. Hence, the additional
contribution coming from the blocking effect can be defined as
\begin{equation}
\delta_{\rm block}=D_{\rm oe}^{\rm pair}-(\Delta+\lambda_2)\;,
\end{equation}
where the $D_{\rm oe}^{\rm pair}$ is the theoretical odd-even difference
of pairing energies. The $D_{\rm oe}^{\rm pair}$ values are calculated using Eq.(1)
with the odd-nucleon blocking effect taken into account. 
Note that blocking also affects the pairing self energy
(Hartree-term), which also
contributes to the $D_{\rm oe}^{\rm pair}$ value.
If the blocking (and deformation changing) effects 
were neglected, we should have $D_{\rm oe}^{\rm pair}\approx\Delta+\lambda_2$.
Since both the $D_{\rm oe}^{\rm pair}$
and $\Delta+\lambda_2$ values increase (decrease) with increasing 
(decreasing) pairing strengths, the $\delta_{\rm block}$ values are 
not very sensitive to the changes in the $G$ values.
We calculate the blocking effects
with the $G$ values obtained by the average gap method \cite{Mol92}.
The results show that the blocking effects are usually about $-200$ to
$-400$ keV for the rare-earth nuclei. 
The shape and blocking effects partially cancel,
but non-zero effects remain systematically.

The obtained shape and blocking effects, $\delta$, are shown in Fig.1.
These values range mostly from $-100$ to $-300$ keV or about 10\% to 30\%
of the corresponding odd-even mass differences, clearly suggesting that
one cannot neglect this component. Note, that the size and the 
fluctuations of $\delta$ reflect two different ways to calculate
odd-even mass differences, and not the difference between experimental
and theoretical odd-even masses.

In general, the shape and blocking effects
change smoothly with particle number and
one would like to separate the contributions from
$\delta_{\rm shape}$ and $\delta_{\rm block}$. 
However, the situation can become rather complex:
For $N=98$--102, the calculated PES's show that the nuclei are soft in
$\beta_2$ deformation, particularly for $^{172-176}$W.
The $\beta_2$ softness results in relatively large uncertainties in
the determination of the $\beta_2$ values and hence
significantly influences the $\delta_{\rm shape}$ values. The same holds
for the $\delta_{\rm block}$ values, leading to fluctuating results.
Hence, separate consideration of the $\delta_{\rm shape}$ and
$\delta_{\rm block}$ values can be misleading. In contrast, the combined
value of the shape and blocking effects is less shape dependent, since
the total energy of a nucleus is not so sensitive
to the deformation value in a range around the minimum of a soft PES.

With the above shape and blocking effect ($\delta$) theoretical
odd-even mass differences ($D_{\rm oe}^{\rm th}$) can be determined,
and compared with experimental values ($D_{\rm oe}^{\rm expt}$) 
to obtain the pairing strengths. The modified $D_{\rm oe}^{\rm th}$
value can be written as
\begin{equation}
D_{\rm oe}^{\rm th}=\Delta+\lambda_2+\delta \;.
\end{equation}
In the practical calculations,
the contribution from quadrupole pairing is included, though this term
is not written explicitly in the above equation. However,
as mentioned above, the contribution of the doubly-stretched 
quadrupole-pairing energies is very small (usually less than 30 keV
in magnitude). In the right side of Eq.(3) the pairing gap, $\Delta$,
is the dominant term.
Obviously, due to the presence of the $\delta$ term, $D_{\rm oe}^{\rm th}$ is
in general not equivalent to the pairing gap ($\Delta+\lambda_2$)
of even-even nuclei.
The $\Delta$ value is very sensitive to the change in the $G$ value,
while the $\lambda_2$ and $\delta$ values are not.

The presence of the negative $\delta$ value implies that
the pairing strength $G$, needs to increase when one aims at
self consistent calculations of odd-even mass differences.
By adjusting the pairing strength, one can reproduce
the experimental $D_{\rm oe}$ value. 
However, in that case we found that each nucleus
requires its separate determination of $G$.
Apparently, the average gap method \cite{Mol92} does not have the proper
particle-number and deformation dependence.
Of course, other 
quantities contributing to the odd-even mass difference
may be lacking in our model. One such effect is the coupling
to phonons, which will influence the ground-state
binding energy, depending on the softness of 
the nuclear shape. Also, displacements of the single-particle spectrum
of the Woods-Saxon potential will affect the calculated $D_{\rm oe}$ values.
To disentangle the different contributions, especially to
optimize a method to determine the average pairing
strength, is outside the scope of the present work.

\epsfysize%
%
%EndExpansion
=12.9 cm 
\centerline{\epsffile{fig1-moi.ps}}%
\begin{figure}[h]
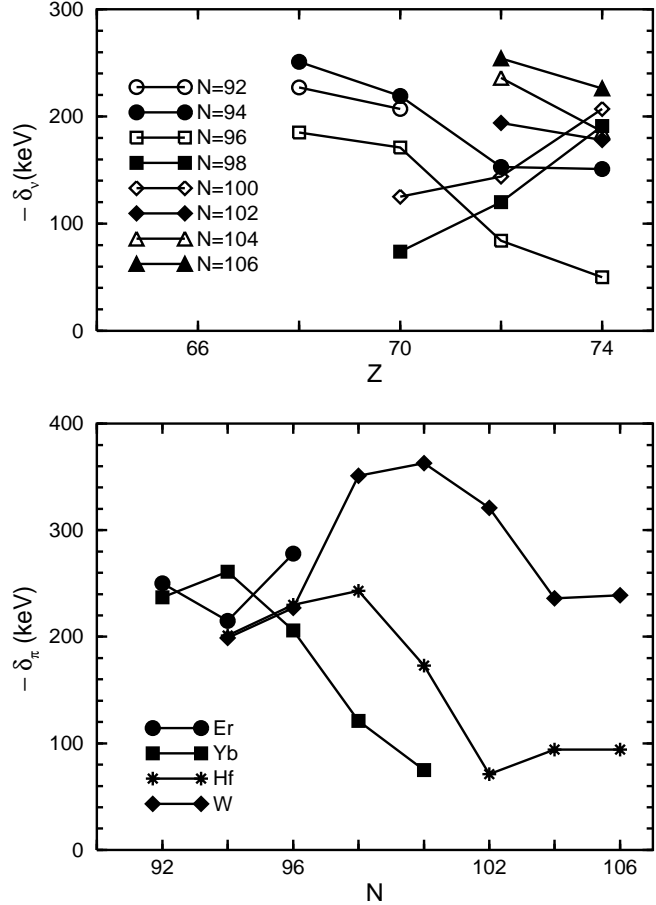

\caption{Obtained shape and blocking effects
($\delta=\delta_{\rm shape}+\delta_{\rm block}$) as a function of
nucleon number. The upper panel is for neutrons ($\nu$) and the
lower panel for protons ($\pi$). Note that the $\delta$ 
values are negative.}
\label{Fig.1}
\end{figure}

In order to better reproduce the average of experimental $D_{\rm oe}$ values,
we scale the pairing strength by $G = F G^0$, where $G^0$ is the pairing
strength obtained by the average gap method \cite{Mol92}.
For reason of simplicity, we use a constant factor $F$,
which we determine from individual 
$F$ values that have been
fitted to the corresponding $D_{\rm oe}^{\rm expt}$ values in this mass region.
%One can consider the adjusting factor, $F$, to be a function of, 
%for example, the nucleon number and deformation.
%However, we assume for the time being that $G^0$ has a reasonable
%$A$-dependence
%for pairing gaps, which on average agree with experimental odd-even
%mass differences \cite{Mol92}, and hence a constant $F$ has been used
%for the local mass region of the studied nuclei.
%The $F$ value can be determined by fitting the $D_{\rm oe}^{\rm th}$ values of Eq.(3)
%to experimental odd-even mass differences ($D_{\rm oe}^{\rm expt}$).
%A simple equivalent method is to calculate an average value among
%the studied nuclei, after individual $F$ values for each nucleus have been
%determined by reproducing the corresponding $D_{\rm oe}^{\rm expt}$ values.
We expect that using an average $F$ value will reduce
the fluctuations arising from the uncertainties of experimental masses and
from possible discrepancies between theoretical and experimental
single-particle levels. For the region of the studied nuclei,
we obtain $\tilde F_\nu=1.08$ (neutrons) and $\tilde F_\pi=1.05$ (protons)
using the experimental masses of Ref.\cite{Aud97}.
This results in increases of the LN pairing gaps by about 25\%
for neutrons and 15\% for protons.

To investigate the consistency of our method,
we calculate the moments of inertia ($J(\omega)$) of yrast rotational bands
by means of the pairing-deformation self-consistent cranked shell model
\cite{Sat94,Sat95}. As mentioned at the beginning of the paper, the moment of
inertia is a very sensitive probe of pairing correlations. 
It is not at all obvious, that a pairing
interaction that reproduces the odd even mass difference at the same
time also can reproduce the moments of inertia.
In Fig.~2, we compare experimentally
deduced moments of inertia with the results of our calculations, done with
the standard pairing strength $G^0$,  and the adjusted 
$G = F G^0$.
Clearly, 
the adjusted $G$ values lead to improved
descriptions of both moments of inertia and  backbending
frequencies. (No additional 
parameters are adjusted in the present calculations.)

\epsfysize%
%
%EndExpansion
=15.9 cm 
\centerline{\epsffile{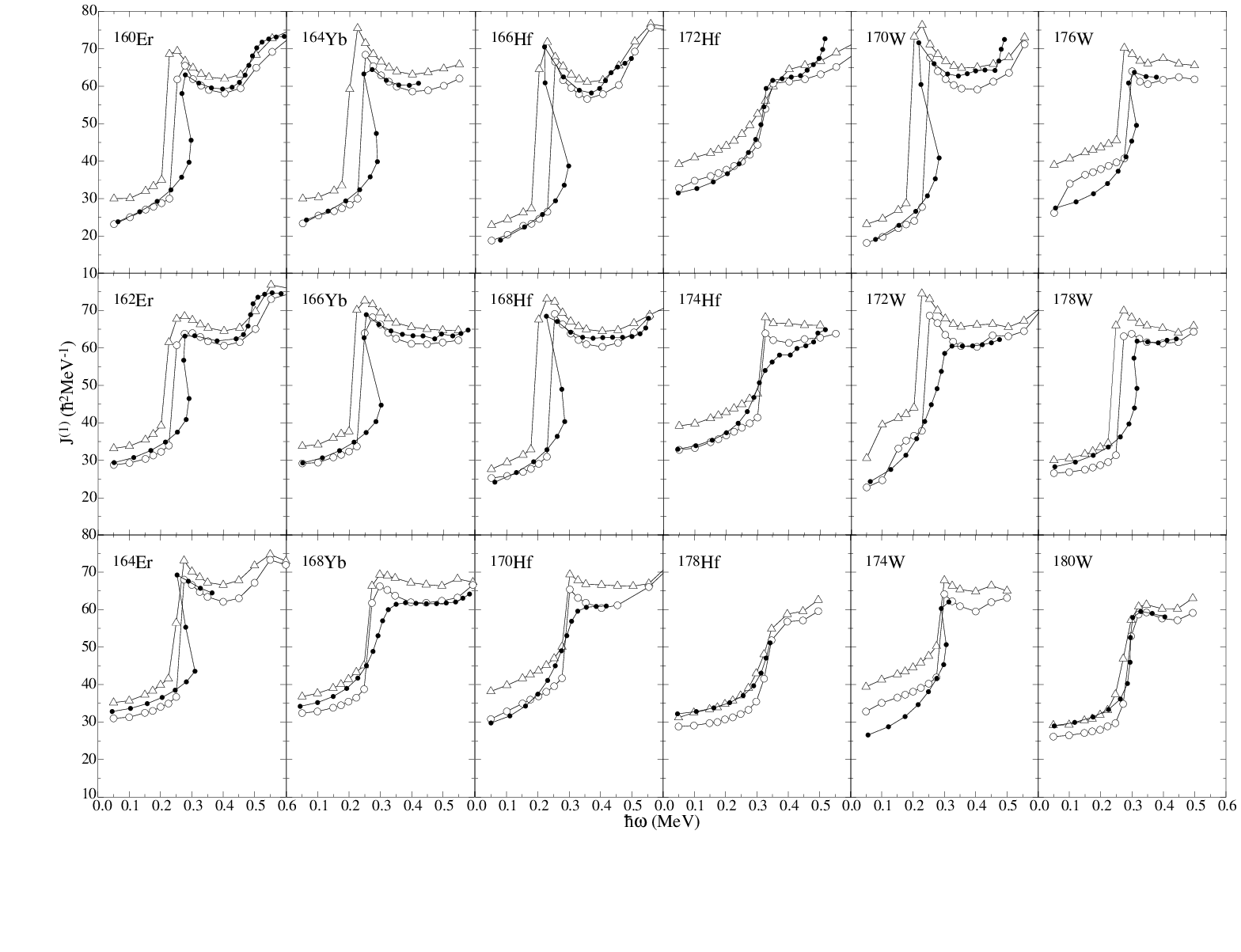}}%
\begin{figure}
%\rotate[r]{}
\caption[]{Calculated and experimental moments of inertia.
The open triangles and circles denote calculations with 
pairing strengths obtained by the average gap method ($G^0$),
and adjusted for average shape and blocking effects
($G={\tilde F} G^0$) respectively.
The dots show the experimental values \cite{Fir96}.}
\label{Fig.2}
\end{figure}

In this context, one needs to recall of the long standing problem
of cranking calculations  with monopole pairing, that
do not at the same time describe both moments of inertia and
band-crossing frequencies (see e.g. \cite{RBen83,TBen89}).
In order to reproduce moments of inertia, one in general needs to use a
{\sl reduced} pairing strength \cite{TBen89,Goe90}.
But on the other hand, an {\sl enhanced} pairing field
is required to reproduce band-crossing frequencies \cite{RBen83}.
The presence of the
time-odd component of the quadrupole pairing field \cite{Die84,Wys95}, 
results in a stiffer nucleus, which allows 
an increase of the $G$ value. Apparently, the doubly stretched
quadrupole pairing interaction in combination with the Lipkin-Nogami
method enables a consistent discription of both band crossing frequencies
and moments of inertia.\footnote{Other effects like the coupling to the
vibrational phonons, that may affect the crossing frequencies
are of course not taken into account}

\epsfysize%
%
%EndExpansion
=9.9 cm 
\centerline{\epsffile{fig3-moi.ps}}%
\begin{figure}
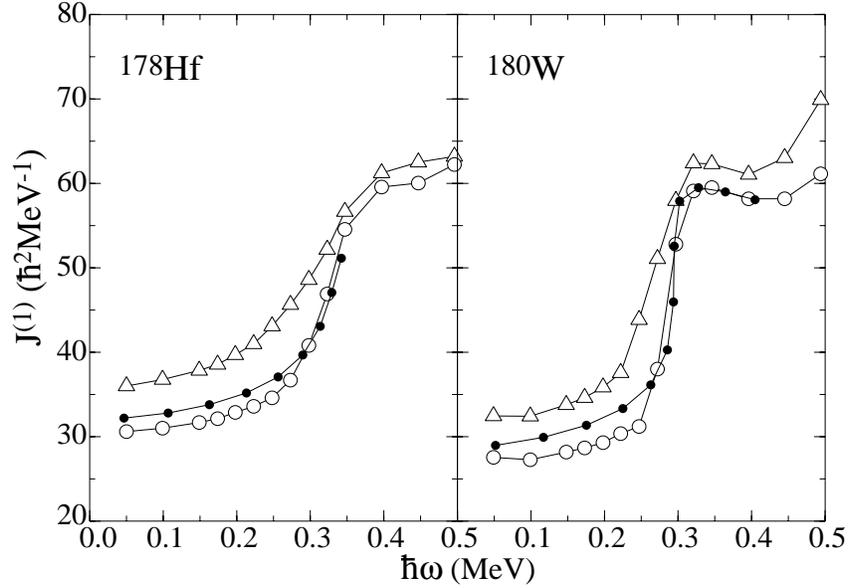

\label{Fig.3}
\caption{Similar to Fig.2, but with the non-average $G$ values.
The open triangles are for the calculations with the $G_0$ values which
reproduce the $D_{\rm oe}^{\rm expt}$ values with $\Delta+\lambda_2$.
The open circles show the results with $G/G_0=1.13$ (neutons)
and 1.05 (protons) in $^{178}$Hf, and correspondingly 1.10 and
1.09 in $^{180}$W, that reproduce the $D_{\rm oe}^{\rm expt}$ values
with $\Delta+\lambda_2+\delta$.}
\end{figure}

For some heavy isotopes, however, using the average $F$ values results in
too small moments of inertia, e.g. in $^{178}$Hf and $^{180}$W.
For these nuclei, the average $F$ values give too large $D_{\rm oe}^{\rm th}$ 
values.
In fact, the $\Delta+\lambda_2$ values obtained from $G^0$ have
already over-estimated the experimental odd-even mass differences
for some heavy isotopes, e.g. by 166 keV (neutrons)
and 190 keV (protons) in $^{178}$Hf and correspondingly
98 and 125 keV in $^{180}$W. In general,
the average gap method gives too large
$\Delta+\lambda_2$ values for heavy isotopes and
too small $\Delta+\lambda_2$ values for light isotopes,
indicating the problems of the A-dependence of the average gap method.
Obviously, averaging the $G$ adjusting factors does not change
the $A$-dependence of the pairing gaps.

In order to check the influences from the possible
discrepancy of the $A$-dependence of pairing gaps,
we have also done the calculation with a pairing strength ($G_0$) 
that reproduces the $D_{\rm oe}^{\rm expt}$ value with $\Delta+\lambda_2$
for each given nucleus. The $G_0$ values are normally smaller than the
average-gap-method $G^0$ values for the heavy nuclei, e.g.
$^{178}$Hf and $^{180}$W.
Results show that the calculated moments of inertia with
such $G_0$ values are systematically larger than the corresponding
experimental values. However, when instead the $G_0$ values
are adjusted by reproducing the $D_{\rm oe}^{\rm expt}$ values with
$\Delta+\lambda_2+\delta$
(i.e. including shape and blocking effects, 
$D_{\rm oe}^{\rm th}=D_{\rm oe}^{\rm expt}$) 
a significantly improved description can be
obtained, as shown in Fig.3 for the $^{178}$Hf and $^{180}$W examples.
Here, we have obtained different $F$ values (see Fig.3 caption)
compared to the above average values,
mainly because the different pairing
strengths ($G^0$ or $G_0$) have been chosen as the reference of the $G$
adjustment. The non-average $F$ values are mostly in range of
1.05--1.10 for neutrons and 1.03--1.08 for protons.
Clearly, the proper pairing strength for odd-even masse differences
is also consistent with experimental moments of inertia.
In addition, the increase of the pairing strength found in our work 
agrees with that needed to reproduce 
the excitation energies of high-seniority states\cite{Xu98}.

Deformations, which can change with rotational
frequency, are determined self-consistently by calculating
the Total Routhian Surfaces (see e.g. \cite{Wys95,Sat94,Sat95}).
With the determined deformations, the calculated intrinsic quadrupole moments
($Q_0$ \cite{Xu98}) agree
with corresponding
experimental values \cite{Ram87}.
The deformation changes due to the adjustments of 
the $G$ values are very small ($|\Delta\beta_2|<0.003$ and
$|\Delta\beta_4|\leq 0.002$) for nuclei that are not soft.
In $^{172,176}$W, some shifts in $J(\omega)$ can
be seen, which are due to the shifts of the $\beta_2$ values with increasing
rotational frequency. The PES calculations, as mentioned,
for $^{172-176}$W are soft in $\beta_2$.

In summary, we have investigated the shape and blocking effects
on odd-even mass differences for even-even
rare-earth nuclei. These effects are shown to be in the range
of 10--30\% of the corresponding odd-even mass differences. 
The blocking effect, in principle, should belong to the category of
pairing effects. 
Clearly, these effects should not be neglected in
determining the pairing strengths. 
Indeed, when blocking and shape effects are taken into account,
pairing strengths are increased by about 5--10\% resulting in sizeble
changes of the pair gaps.
The adjusted strengths are consistent with what is needed
to reproduce the excitation
energies of multi-quasi particle configurations,
and lead to an improved description of nuclear collective
rotational motion, through calculating moments of inertia and backbending
frequencies. 
The present work establishes a consistent relation between mass differences,
moments of inertia and excitation energies of high seniority states.

This work was supported by the UK Engineering and Physical Sciences
Research Council and the Swedish Natural Sciences Research Council.

\end{document}